# Electro-Optically Tunable Universal Metasurfaces

Ghazaleh Kafaie Shirmanesh[1], Ruzan Sokhoyan[1], Pin Chieh Wu[1,2], and Harry A. Atwater[1,3*]

[1]Thomas J. Watson Laboratory of Applied Physics, California Institute of Technology, Pasadena, California 91125, USA
[2]Department of Photonics, National Cheng Kung University, Tainan 70101, Taiwan
[3]Kavli Nanoscience Institute, California Institute of Technology, Pasadena, California 91125, USA

*Corresponding author: haa@caltech.edu

**Molding the flow of light at the nanoscale has been a grand challenge of nanophotonics for decades. It is now widely recognized that metasurfaces represent a chip-scale nanophotonics array technology capable of comprehensively controlling the wavefront of light via appropriately configuring subwavelength antenna elements. Here, we demonstrate a reconfigurable metasurface that is universal, i.e., notionally capable of providing diverse optical functions in the telecommunication wavelength regime, using a compact, lightweight, electronically-controlled array with no moving parts. By electro-optical control of the phase of the scattered light from identical individual metasurface elements, we demonstrate a single prototype universal programmable metasurface that is capable of both dynamic beam steering and reconfigurable light focusing using one single device. Reconfigurable universal metasurfaces with arrays of tunable optical antennas thus can perform arbitrary optical functions by programmable array-level control of scattered light phase, amplitude, and polarization, similar to dynamic and programmable memories in electronics.**

R apid advances in control of the phase and amplitude of the light scattered from planar arrays of nanophotonic elements has stimulated the development of metasurfaces that utilize amplitude/phase-sensitive scattering to enable wavefront engineering[1, 2]. Metasurfaces are also now demonstrating some of their potential applications in compact, high-performance, and low-cost optical devices and components, creating burgeoning interest in photonic integration. To date, metasurfaces have mostly been designed in an application-specific manner and the design process resulted in bespoke architectures tailored to particular applications. Dynamical control of the properties of the scattered light is possible by using tunable metasurfaces, for which external stimuli such as electrical biasing, optical pumping, heating, or elastic strain can give rise to changes in the dielectric function or physical dimensions of the metasurface elements[3], thereby modulating the antenna phase and amplitude response. Among these mechanisms, electrical tuning has been proven to be a robust, high speed, energy-efficient and reversible scheme for tuning active metasurfaces[4-11].

The ability of metasurfaces to spectrally, temporally, or spatially manipulate the wavefront of light with very high spatial resolution, is expected to accelerate miniaturization of optical devices and integration of optical systems[12]. However, in spite of advances in active metasurfaces to date, universally reprogrammable metasurface components have not yet been demonstrated. Realization of a single hardware device that can provide multiple and --indeed general-- functions would further accelerate the impact of metasurfaces and their applications. Such universality can be found in electronics technology that has benefitted from development of programmable and reprogrammable circuits composed of identical circuit elements, such as dynamic[13] and static[14] random access memories and field-programmable gate arrays[15]. In this paper, we demonstrate a state-of-the-art prototypical 'universal' metasurface which can be electronically programmed to achieve two of the most essential functions identified to date for metasurfaces, namely, beam steering and focusing of light.

Optical beam steering is the key element of a broad range of optical systems such as light detection and ranging (LiDAR)[16], optical interconnects[17], and optical communications[18]. Conventional beam steering devices such as Risley prisms[19], galvanometer-scanning mirrors[20], and decentered lenses[21] employ mechanically moving optical components to steer the incident light. Although mechanical beam steering systems provide wide steering angular range and large number of resolvable beam directions, they suffer from low steering speed due to the inertia of their moving parts and the weight of their mechanical components[22]. The availability of electronic beam steering arrays at near infrared (NIR) wavelengths with scanning frequencies above the MHz range could replace mechanical components with compact and lightweight optoelectronic alternatives and enable new functions unachievable via mechanical motion.

Reconfigurable metasurfaces have recently been employed to provide dynamic beam steering in the microwave and NIR regimes by exploiting microfluidic flow[23, 24], incorporation of phase-change materials[25], and reorientation of liquid crystals[26]. However, the performance for these devices is limited by low (~kHz) switching speeds. Alternatively, electro-optic modulation in multiple-quantum-well resonant metasurfaces[27], an intrinsically ultrafast process, has been shown to provide high-speed dynamic beam steering, but to date, a limited phase modulation range has constrained the achievable beam directivity and steering angle range.

Electro-optically controllable beam switching has also been demonstrated via incorporation of transparent conducting oxides as active media into the metasurfaces[4, 28, 29]. However, individual control over each metasurface element, which is required for more complex phase distribution patterns, has not been reported. Other researchers have demonstrated beam steering using waveguide-based thermo-optical phase shifters coupled to antennas[30-34], or by employing frequency-gradient metasurfaces[35]. These chip-based antenna arrays can enable beam steering at visible or infrared frequencies, but are application-specific, and hence, has been unable to achieve more general array functionalities.

Light focusing is another paramount optical function that plays a fundamental role in almost every optical system such as imaging, microscopy, optical data storage, and optical encryption[36]. Metasurfaces have given rise to versatile metalenses that can replace bulky conventional lenses by engineering abrupt phase delays introduced by individual metasurface elements and phase gradients across antenna arrays[37-39]. In order to focus light using metasurfaces, the spatial variation of field amplitude or phase distribution has to be carefully controlled over arrays of resonant elements at approximately wavelength-scale or smaller spacing. Metalenses have demonstrated the capability to perform high-resolution imaging, wavefront shaping for aberration correction, and polarization conversion[1, 2, 40].

Reconfigurable metasurfaces have been utilized to realize dynamic focusing by variation of the overall lens optical thickness or curvature, via liquid crystal reorientation[41], microfluidic flow[42, 43], or elastic deformation[44]. We note these modes of dynamic focusing do not permit precise tailoring of the lens focal properties by arbitrary phase control of the lens elements. Alternatively, electro-thermo-optically controllable lenses have been proposed to precisely engineer the optical wavefronts at the microscale[45]. However, such a dynamic wavefront control at the subwavelength scale, an essential requirement to achieve high resolution imaging, has not yet been reported.

Here, we design and demonstrate a universal electro-optically tunable metasurface that can exhibit multiple functions in the NIR wavelength regime using a single device, via precise tailoring of the phase profile of an optical aperture. Figure 1a schematically illustrates this metasurface, whose independently addressable elements enable dynamic control of the wavefront via a pixel-by-pixel reconfiguration. Using this scheme, we demonstrate a reprogrammable metasurface whose function can be reconfigured between dynamic beam steering and dynamic cylindrical metalens, achieving a reconfigurable focal length and numerical aperture by tuning the gate voltages applied to individual metasurface elements.

Figures 1, b and c schematically illustrate the building blocks of our tunable gated field-effect metasurface, consisting of an Au back-reflector, on top of which an $Al_2O_3$ layer is deposited. The $Al_2O_3$ layer acts as a dielectric spacer, adding a degree of freedom for the metasurface optical mode profile design. This layer is followed by deposition of an indium-tin-oxide (ITO) layer, a gate dielectric, and Au 'fishbone' nanoantennas. The fishbone nanoantennas are comprised of patch antennas that are connected together by Au stripes, which also serve as gate voltage control electrodes. The gate dielectric is a hafnium/aluminum oxide nanolaminate (HAOL), a hybrid material that simultaneously exhibits high breakdown field and high DC permittivity[6]. We apply a DC electric bias between the ITO layer and the nanoantennas. This causes the ITO layer to undergo a reproducible field-effect-induced index change. By altering the applied electric field, one can modulate the ITO charge carrier density close to the interface of the ITO and the gate dielectric. By further increasing the applied bias, the real part of the dielectric permittivity in an accumulation layer located within ITO takes values between -1 and +1, yielding an epsilon-near-zero (ENZ) condition. In the ENZ regime, the ITO layer permittivity is varied at NIR wavelengths by changing the applied DC bias. The width and length of the antenna, and the width of the electrode are designed so that a magnetic dipole plasmon resonance occurs at the wavelengths coinciding with the ENZ regime for ITO, operating in the telecommunication wavelength regime. As a result of the spectral overlap of the ENZ regime of ITO and the geometrical resonance of the metasurface, the metasurface is expected to exhibit large phase modulation.

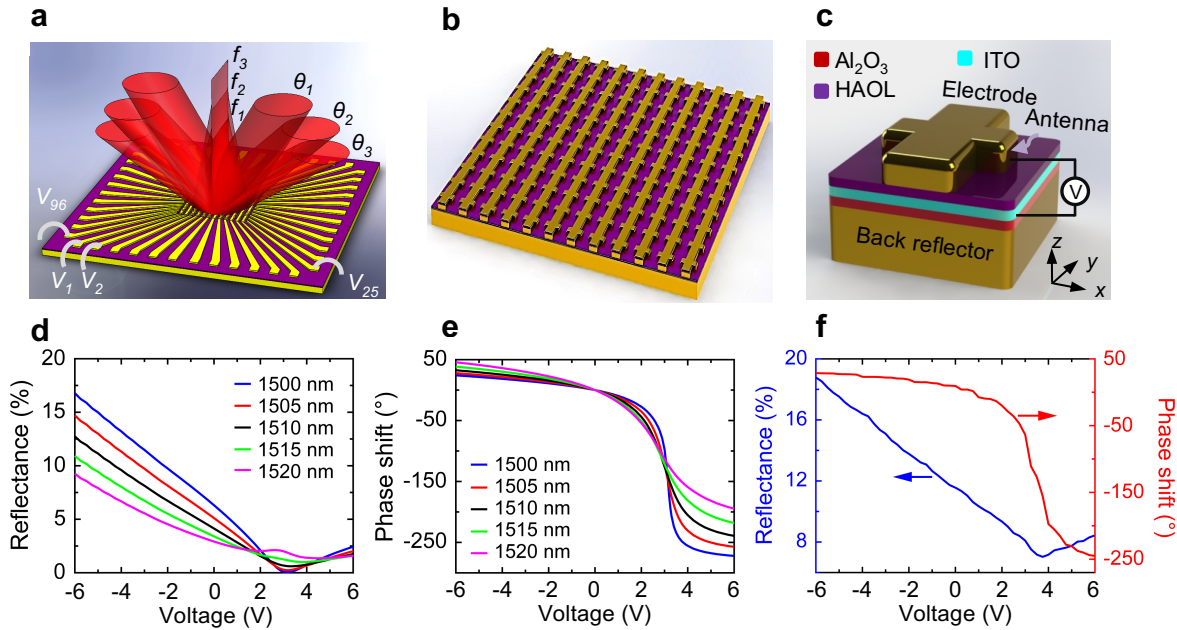

**Figure 1. a,** Schematic of the universal metasurface whose functionality can be switched between dynamic beam steering and cylindrical metalens with reconfigurable focal length. Schematic of (**b**) periodic array and (**c**) unit cell of the antenna elements. The metasurface is composed of an Au back-reflector, an $Al_2O_3$ dielectric layer, an ITO layer, and a hafnium oxide/aluminum oxide laminated (HAOL) gate dielectric

followed by an Au fishbone antenna. The periodicity of the metasurface is 400 nm, and the thickness of the back-reflector, $Al_2O_3$, ITO, and HAOL layers are 80 nm, 9.5 nm, 5 nm, and 9.5 nm, respectively. The width, length, and the thickness of the antenna are 130 nm, 230 nm, and 40 nm, respectively and the width of the electrode is 150 nm. Simulated (**d**) reflectance and (**e**) phase of the reflection from the metasurface as a function of applied voltage for different wavelengths. **f,** Measured reflectance (blue curve) and phase shift (red curve) as a function of applied bias voltage. The operation wavelength of the fabricated device was chosen to be $\lambda$=1522 nm such that a phase shift of greater than 270°, accompanied by a modest reflectance change could be obtained.

Figures 1, d and e, respectively, illustrate the simulated reflectance and phase shift as a function of applied bias at different wavelengths. Here, phase shift is defined as a difference between the phases of the reflected and incident plane waves calculated at the same spatial point. As can be seen, when the external bias is changed, we observe a reflectance change that is accompanied by significant phase modulation. This demonstrates that both the real and imaginary parts of the refractive index of the active region in the ITO layer are modulated by the applied bias. Once we obtained the reflectance and phase shift spectrum of the designed metasurface under applied bias, we can then pick the operation wavelength of the beam steering and focusing devices. To accomplish this, we utilize the metasurface as a phase modulator where the reflectance should ideally remain constant upon change in the applied bias. Here the operation wavelength of $\lambda$=1510 nm is chosen so that we obtain a phase shift of higher than 270° while the maximum reflectance modulation remains as modest as possible. After confirming this tunable response, we experimentally obtained the reflectance and phase shift of the fabricated metasurface under applied bias. Figure 1f illustrates the measured reflectance (blue curve) and phase shift (red curve) as a function of applied bias. In order to experimentally evaluate the reflection phase shift from the metasurface, we used a Michelson interferometer system[6, 8]. By focusing the incident laser beam on the edge of the metasurface nanoantenna array, the scattered beam is reflected partly from the metasurface and partly from the gold back-plane, resulting in a lateral shift in the interference fringe patterns of the metasurface and the back-reflector when changing the applied bias. By fitting these two cross sections to sinusoidal functions, and obtaining the relative delay between the fitted sinusoidal curves when changing the applied voltage, we could retrieve the phase shift acquired due to the applied bias. As can be seen in Fig. 1f, a significant phase shift of 273.81° accompanied by a modest reflectance modulation is attainable at the device operation wavelength by electrically biasing the metasurface.

Once we validated the modulation performance of the individual metasurface elements, we investigated the metasurface array beam steering and focusing performance. Scanning electron microscopy (SEM) images of the fabricated metasurface nanoantennas are shown in Fig. 2a. In our metasurface device, nanoantennas are electrically bus-connected together in one direction, forming equipotential antenna rows, referred to here as a metasurface pixel. Then each pixel is individually controlled by a separate applied gate voltage. Figure 2b is a photomicrograph of the fabricated array, consisting of 96 individually-controllable and identical metasurface pixels. In order to individually bias each of these metasurface pixels, we designed two printed circuit boards (PCBs). The sample is mounted on the first PCB ($P_1$), and 96 individual metasurface elements as well as 4 ITO connecting pads (to be used as the ground) are wire-bonded to 100 conducting pads on the PCB. Figure 2c shows the first PCB with the multifunctional metasurface mounted on and wire-bonded to it. Each conducting pad on $P_1$ is then connected to an external pin on the second PCB ($P_2$) that is shown in Fig. 2d. This voltage deriving PCB is capable of providing 100 independent voltages that can be individually controlled through programming a number of micro-controllers by a computer.

Then, in order to characterize our universal metasurface, we used a custom-built optical setup that could measure reflectance spectrum, phase shift, beam steering profile, and focused beam profile.

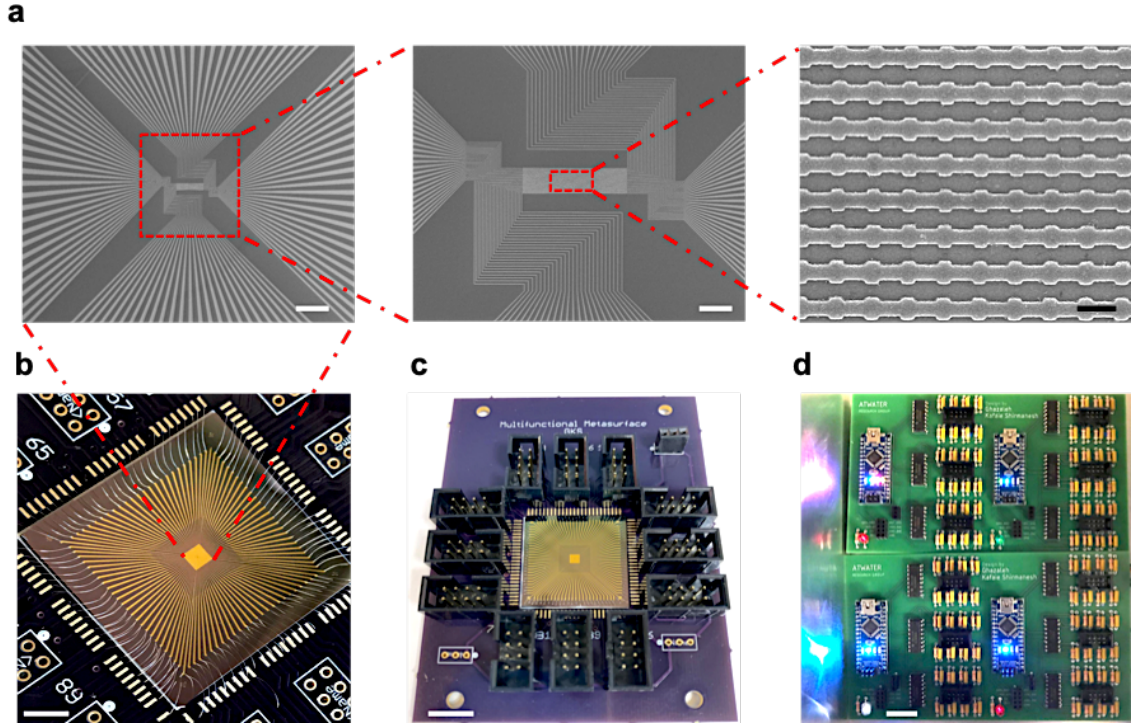

**Figure 2. a**, SEM image of the nanoantennas of the fabricated gate-tunable metasurface for the demonstration of dynamic beam steering and a reconfigurable metalens. The scale bars from left to right are 200 μm, 50 μm, and 500 nm respectively. **b**, Photographic image of the universal metasurface with 96 independently addressable elements. Scale bar is 5 mm. **c**, Sample mounting PCB to which we wire-bond the universal metasurface pads ($P_1$). 96 metasurface elements' pads and 4 ITO pads are wire-bonded from the sample to 100 conducting pads on $P_1$. Scale bar is 10 mm. **d**, Voltage deriving PCB ($P_2$) that provides 100 voltages controlled by programming micro-controllers. Scale bar is 20 mm.

**Demonstration of Beam Steering**

After validating the wide phase tunability of our metasurface, we designed and demonstrated a dynamic beam steering device. In order to implement beam steering, we used a blazed grating approach to design the spatial profile of the phase of the light reflected from the metasurface by engineering the spatial distribution of the DC bias voltages applied to the 96 metasurface pixels.

When no bias was applied, we observed only the zeroth order diffracted beam in the Fourier plane. In other words, the subwavelength period of the metasurface results in an absence of higher-order diffracted beams at zero bias. Then we designed the spatial phase profile of the beam steering metasurface by discretizing the phase shift acquired by the metasurface pixels into four levels 0°, 90°, 180°, and 270°. Each effective period, hereafter termed a supercell, consists of the metasurface pixels exhibiting the discretized 4-level phase shift values. By changing the pixel repetition number (RN) for each phase shift value within one supercell, we electrically modulated the effective periodicity of the metasurface array. This resulted in a shift of the spatial position of the first diffracted order, enabling manipulation of the far-field radiation.

Figure 3a shows the metasurface spatial phase profiles, for the four-level phase shift with different RN values. In Fig. 3a, each gray-shaded region determines one supercell in each case. The simulated far-field pattern of the beam steering device is presented in Fig. 3b. As can be seen, by changing the RN value, the size of the metasurface supercell is electrically modulated, resulting in reconfigurable beam steering with quasi-continuous steering angles that can be as large as ~70.5°. Figure 3c shows the measured far-field pattern for our fabricated beam steering device whose SEM images showed an average pitch size of 504 nm.

Due to limitations of our measurement setup, steering angles of higher than 23.5° could not be captured by the imaging system. As a result, the maximum measured steering angle was ~22°, which corresponds to a repeat number of 2. As expected, by increasing the effective period of the metasurface, the beam angle becomes smaller. We also note that for each RN value, no diffracted order is observed at negative angles, indicating true phase gradient beam steering rather than switchable diffraction. This confirms that the beam steering is obtained as a result of the asymmetric phase gradient introduced by the subwavelength metasurface phase elements.

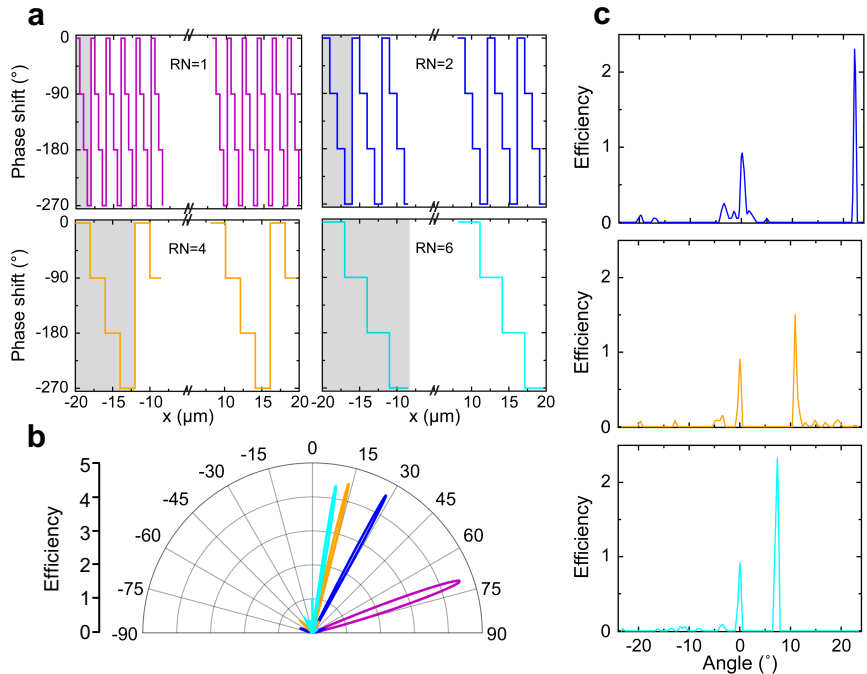

**Figure 3. a**, The spatial phase distributions of the metasurface elements with different RN values that are used to create phase gradients resulting in beam steering. **b**, Simulation results of the beam steering metasurface obtained through analytical calculations. Changing the RN value, the steering angles of 70.68° (RN=1), 28.14° (RN=2), 13.62° (RN=4), and 9.02° (RN=6) were obtained through calculations. **c**, Experimental results of the beam steering metasurface. Changing the RN value, we could obtain the steering angles of 22.19° (RN=2), 10.91° (RN=4), and 7.40° (RN=6). Each steering angle corresponds to the spatial phase distribution of the same color presented in **a**.

**Demonstration of Dynamic Metalens**

Using the same concept of controlling the phase imposed by each individual metasurface pixel, we were able to demonstrate the use of our universal metasurface as a reconfigurable lens by developing phase profiles for lenses with different focal lengths. Figures 4, a-c show the spatial distribution of the phase shift (diamond) and the corresponding applied bias voltage (square) required to focus the reflected beam at focal lengths of 1.5 μm, 2 μm and 3 μm, respectively. These values were extracted from the simulated phase shift as a function of applied bias. In order to investigate the focusing performance, we simulated the universal metasurface under the applied bias distributions illustrated in Figs. 4, a-c. In our full-wave electromagnetic simulations, we modeled a miniaturized lens with a 20 μm aperture size since simulating the full metasurface at the small mesh sizes required for the ITO layer active region is beyond our present numerical simulation capability. Figures 4, d-f illustrate the far-field pattern of the beam reflected from our tunable metasurface in the x-z plane. As seen in Figs. 4, d-f, the metasurface can clearly focus the reflected light at the focal lengths of 1.5 μm, 2 μm and 3 μm, when appropriate bias voltages are applied to the individual metasurface pixels.

We then experimentally characterized metasurface dynamic focusing once the focusing performance of our universal metasurface was confirmed by calculations. We programmed the voltages applied to each metasurface pixel in order to experimentally achieve the desired phase shift values (Fig. 1f). Then the fabricated metalens was characterized utilizing our universal setup. Using this setup, the intensity profile of the reflected beam in the *xy*-plane was recorded. By extracting the cross sections of the captured intensity profiles at fixed *y* values, we reconstructed the intensity profile of the reflected beam in the *xz*-plane. Figures 4, g-i illustrate the metalens reflected beam intensity profiles in the *xz*-plane for the applied bias distributions shown in Figs. 4, a-c. As can be seen, the fabricated metasurface focuses the reflected beam at the desired depths. The scale bars in Figs. 4, g-i were obtained by imaging an object of known size. When the incident light was polarized perpendicular to the antennas, no focusing was observed since no phase modulation could be achieved in that polarization. This observation confirmed that the captured focusing originated from the metasurface.

In conclusion, we have designed and experimentally demonstrated an electrically-tunable universal metasurface in the NIR wavelength range. The universal metasurface is realized via field-effect-induced modulation of transparent conducting oxide active regions incorporated into the metasurface, and is capable of spatiotemporal modulation of the fundamental attributes of light. As a proof of concept, we designed phase profiles for our universal metasurface to demonstrate beam steering and dynamic focusing using the same device via individually controlling each metasurface pixel. Such a universal metasurface can initiate integrated on-chip electro-optical devices such as light detection and ranging (LiDAR) systems. A worthy direction for future research is to extend the universal metasurface concept demonstrated here to a two-dimensional phased array architecture. In addition to enabling beam steering and focusing in two dimensions, such a two-dimensional array could enable fast and energy-efficient programmable devices such as dynamic holograms, off-axis lenses, axicons, vortex plates, and polarimeters.

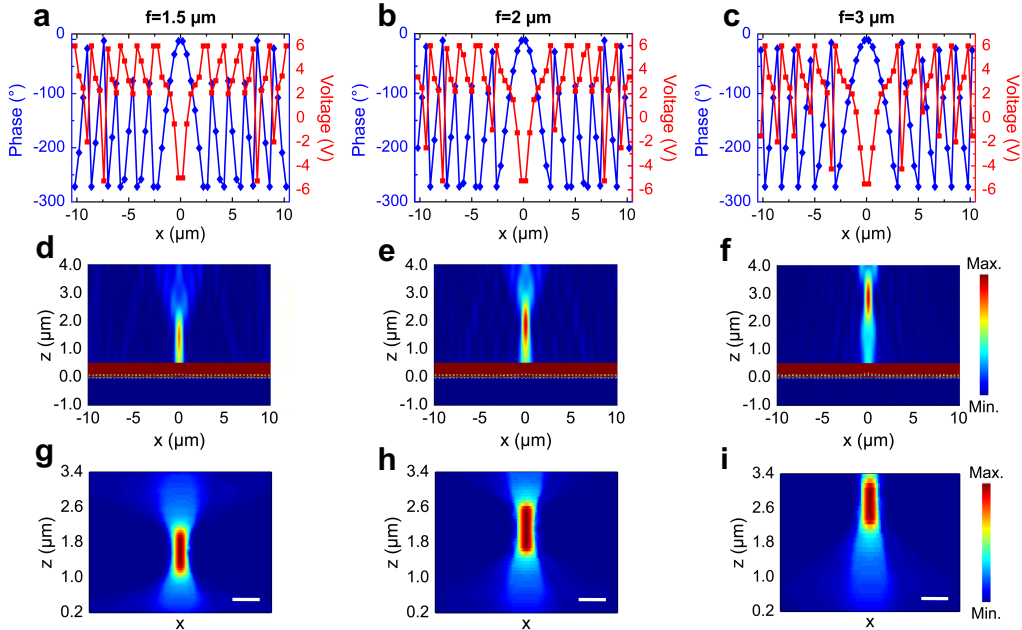

**Figure 4.** Spatial phase (diamond) and voltage (square) distribution of a metalens with focal lengths of (**a**) f=1.5 μm, (**b**) f=2 μm, and (**c**) f=3 μm using the phase shifts obtained from the simulation. Full-wave simulation of the spatial distribution of the electric field |E|² for the metalens with focal lengths of (**d**) f=1.5 μm, (**e**) f=2 μm, and (**f**) f=3 μm. Measured intensity profile of the beam reflected from the metalens with focal lengths of (**g**) f=1.5 μm, (**h**) f=2 μm, and (**i**) f=3 μm. Scale bar is 2 μm.

**Acknowledgements**

The authors acknowledge metasurface device fabrication support provided by the Kavli Nanoscience Institute (KNI).
**Funding:** This work was supported by Samsung Electronics and the National Aeronautics and Space Administration. P.C.W. acknowledges the support from Ministry of Science and Technology, Taiwan (Grant number: 107-2923-M-006-004-MY3; 108-2112-M-006-021-MY3). P.C.W. also acknowledges the support in part by Higher Education Sprout Project, Ministry of Education to the Headquarters of University Advancement at National Cheng Kung University (NCKU).